\documentclass[twocolumn,showpacs]{revtex4}

\usepackage{graphicx}
\usepackage{bm}

\begin{document}

\title{Screening effects on $^1S_0$ pairing in neutron matter}

\author{Caiwan Shen$^{1,2}$, U. Lombardo$^{1,3}$, and P. Schuck$^{4}$, 
        W. Zuo$^{5}$, N. Sandulescu$^{6}$}
\affiliation{
     $^{1}$ INFN-LNS, S. Sofia 44, I-95123 Catania, Italy \\
     $^{2}$ China Institute of Atomic Energy, P.O.Box 275(18), Beijing 102413, China \\
     $^{3}$ Dipartimento di Fisica, Via S. Sofia 64, I-95123 Catania, Italy \\
     $^{4}$ Institut de Physique Nucl\'{e}aire, Universit\'{e} Paris-Sud, 
            F-91406 Orsay Cedex, France \\
     $^{5}$ Institute of Modern Physics, Lanzhou 730000, China \\
     $^{6}$ Institute of Atomic Physics, P.O.Box MG-6, Bucharest, Romania
          }

\begin{abstract}
The $^1S_0$ superfluidity of neutron matter is studied in the framework of
the generalized Gorkov equation. The vertex corrections to the pairing
interaction and the self-energy corrections are introduced and 
approximated on the same footing in the gap equation. A suppression of the 
pairing gap by more than $50\%$ with respect to the BCS prediction is
found, which deeply changes the scenario for the dynamical and
thermal evolution of neutron stars. 
\end{abstract}


\maketitle


Neutron superfluidity in neutron star matter though an old subject, is still of great
actuality and vividly debated (for a review, see \cite{SCHUL}). 
The reason for this stems from the  fact that 
superfluidity is an extraordinarily subtle  process when it comes to quantitative 
predictions starting from the bare NN interaction. On the other hand for 
neutron stars, such quantitative predictions are necessary, since the manifestation of
the superfluidity are only rather indirect through glitches 
and relaxation phenomena and cooling rates.
One, therefore, lacks direct experimental information on the magnitude of neutron pairing. 
On the other hand, there is no doubt that the dynamics 
and the thermodynamics of neutron stars are strongly 
influenced by the superfluid character of neutron matter and it is therefore important 
to get pairing properties of neutron-star matter under better control from the microscopic
point of view. Of course, neutron matter superfluidity is not completely decoupled from 
the one prevailing in finite nuclei. Though we have experimental information for these 
objects, also in this case the fully microscopic explanation of the observed phenomena
is far from being completely settled. It may be argued that in finite microscopic systems,  
where the surface plays a very important role, the situation can be quite different from
the homogenous case. However, like for other quantities of nuclear physics it should be
possible to disentangle volume and surface effects also for pairing properties in finite 
nuclei and it is therefore our belief that the topic of superfluidity in neutron matter, 
nuclear matter and finite nuclei should be studied in an interrelated way. 

    In this work we again concentrate on neutron matter in pursuing previous studies. 
However, it is planned to parallel this work for nuclear matter in the near future. 
In the past we have mainly been concentrating  on the influence of either dynamic self-energy 
corrections \cite{ZUO} (see also \cite{BAL}) {or} vertex corrections 
to the neutron matter pairing problem \cite{CTLI}.
All investigations
in this direction invariably led to the conclusion that dynamic self-energy corrections
yield a quite strong reduction of $^{1}S_{0}$ pairing in neutron matter. However, to be
consistent, self-energy corrections have to be followed by vertex corrections on the 
same footing. This is the objective of the present investigation. It will be seen 
that the vertex corrections have tendency to further reduce the gap but to a lesser 
extent than that is the case from self-energy corrections.  Our approach is based 
on the Gorkov Green's function formalism where we develop systematically self-energy
and effective pairing interaction to lowest order in the particle-hole bubble insertion. 
Though intuitively quite reasonable and physically motivated, we should mention that 
this is neither based on an expansion in a small parameter nor does it rely on some 
variational principle. None the less our results will be quite comparable to those 
of other works, notably to those of Clark et al. \cite{CLARK} who based their investigation 
on the correlated basis function approach which, to a large extent, is variational. 


The superfluid phase of a homogenous system of fermions is described  by 
the pairing field $\Delta_k(\omega)$, which is the solution of the 
generalized gap equation 
\begin{equation}
\Delta_{\vec k}(\omega)=\sum\limits_{k'}\int \frac{d\omega'}{2\pi i}
{\cal V}_{\vec k,\vec k'}
(\omega,\omega') F_{\vec k'}(\omega').
\end{equation}
Here $\cal V$ is the sum of all irreducible nucleon-nucleon (NN) interaction 
terms and $F_k(\omega)$ the anomalous propagator \cite{NOZ,MIG,RING}. 
In most pairing calculations \cite{SCHUL} the effects 
of the medium polarization has been included
in the self-energy \cite{BAL,ZUO}, and not in the pairing potential. A more
general study requires the medium corrections to be treated on equal footing 
also in the vertex corrections as well as in the self-energy. This is the main 
concern of the present letter. Accordingly, the expansion of the interaction 
block $\cal V$ and the self-energy $\Sigma$ have both been truncated to second order
of the interaction. The corresponding diagrams are depicted in Fig. 1. 

The limitation to lowest order bubble insertion may seem as a strong restriction 
because resummation of the bubble series into RPA can have an important influence. 
However, in the exploratory work where self-energy and vertex corrections are 
treated consistently for the first time, we think that higher order effects 
unnecessarily complicate the approach and that the lowest order effects in the 
density should at least give the correct tendency even up to densities around saturation.

The matrix elements of the bare interaction V exhibit hard-core 
divergences which have to be removed by dressing them with the short-range particle-particle
correlations (ladder diagrams). This can be done either in a microscopic approach
by replacing V with the G-matrix or in  a semi-phenomenological approach by replacing 
V with an effective interaction.
   
In the calculations of all diagrams, the superfluid propagators have been replaced by
the normal phase propagators. This amounts to neglecting second-order corrections 
in the gap, which are actually negligible as shown in a fully self-consistent 
calculation discussed below. 

\begin{figure}[htb]
\includegraphics[width=7cm]{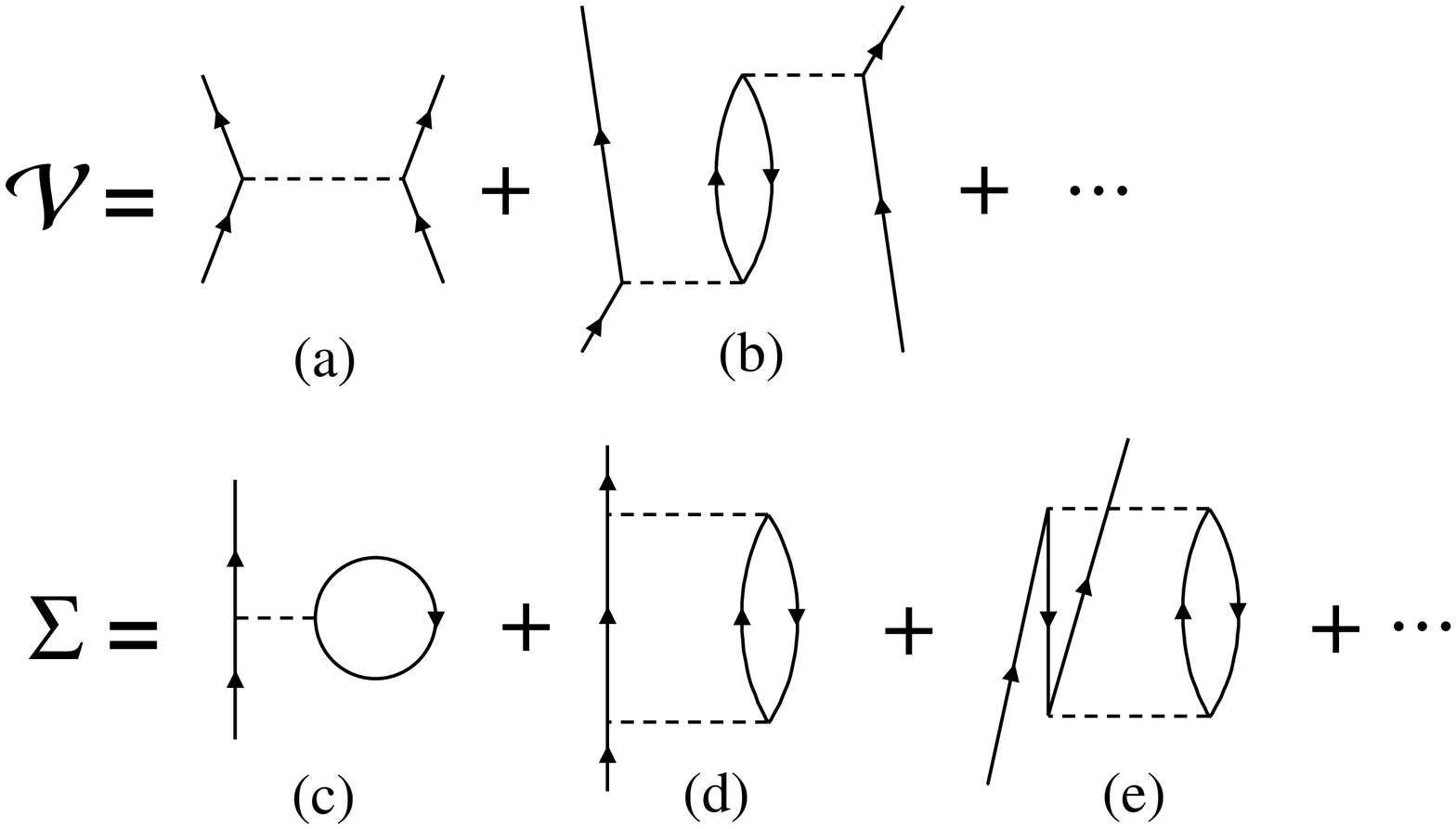}
\caption{\label{fig1} The diagrams of NN interaction and self-energy
discussed in the text. The exchange terms are understood.}
\end{figure}

In general, the interaction as well as the self-energy are complex
quantities which implies the energy gap to be a complex function. The 
introduction of the imaginary part of the self-energy amounts to taking
into account the effects of the quasi-particle spectral function, which have 
been studied elsewhere \cite{BOZ,IMA}.
The complex nature of the potential is due to finite time propagation
and decay of processes shown in Fig. 1. 
In the present investigation $\cal V$ will be assumed to be a real function 
what implies the gap function to be real also.

\begin{figure}[htb]
\includegraphics[width=85mm]{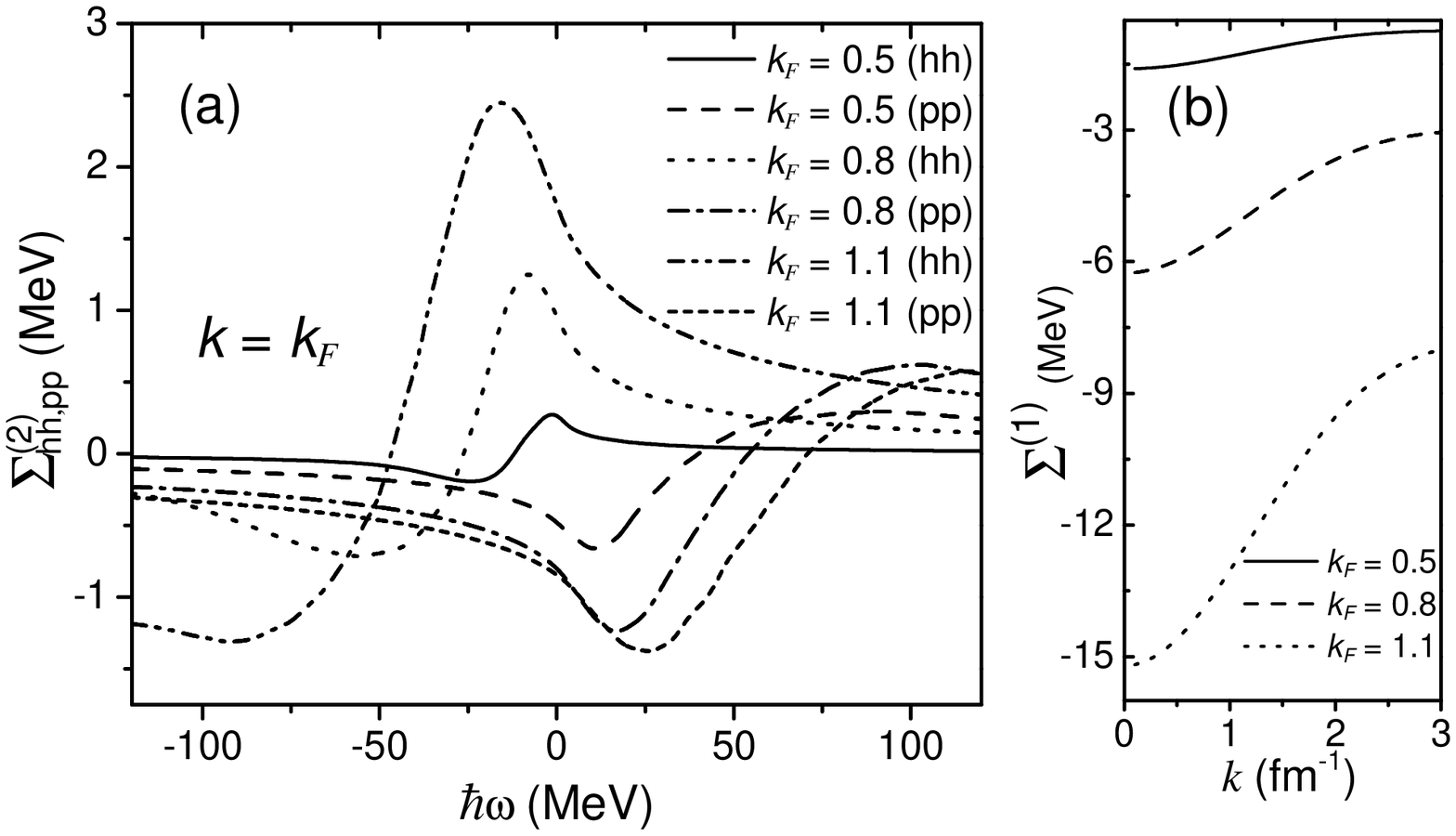}
\caption{ (a) Rearrangement contributions to the self-energy, where $k$ is
         fixed to $k_F$. (b) The HF mean field is plotted vs momentum $k$ at
         $k_F$ = 0.5, 0.8, 1.1 fm$^{-1}$.}
\label{fig2}
\end{figure}

For the present calculation we adopted the semi-phenomenological approach and chose
the Gogny force D1 as effective interaction at each coupling vertex shown in 
Fig. 1. The replacement of the vertices in Fig. 1 by the Gogny force which 
can be considered as a phenomenological representation of a $G$-matrix 
may seem unjustified for the lowest order term in Fig. 1(a), since 
perturbation theory tells us that it should be the bare NN interaction. 
However, it is well known \cite{LOMBA} that limiting the $k$-space, the bare interaction 
has to been replaced by an effective one and choosing the cut off to be at 
$k_F$ the effective force in Fig. 1(a) can be shown to be again equivalent to the 
$G$-matrix. In neutron matter only the density independent part of the Gogny 
force survives with a range in $k$-space of about $k_F$ at the saturation density. 
Therefore, from this point of view, it may not be unreasonable to take the 
Gogny force even for the lowest order term in Fig. 1(a), which is further 
backed by the fact that the gap with the Gogny force 
(without polarization terms) is quite close to the one calculated from a 
bare NN force. In any case the qualitative influence of the polarization 
on the lowest order solution of the gap equation, given by retaining only 
graphs Fig. 1(a) and Fig. 1(c), should not depend very much whether we use in Fig. 1(a) 
the bare interaction or the Gogny force.

 The self-energy $\Sigma_k(\omega)$ in neutron matter has
been calculated within the above described approximation. The first-order 
contribution is reported in Fig. 2(b) for three values of the nuclear 
matter density in the range where the pairing is expected to 
be largest. The second-order contributions 
$\Sigma^{(2)}_{pp}$ and $\Sigma^{(2)}_{hh}$ (graph b) and c)) in Fig. 1, respectively)
are depicted in Fig. 2(a) as a function of the energy. Only the energy dependence of
of these terms will be discussed, since the Gogny force implicitly contains 
the static part, which will be removed from the gap equation as described later on.

$\Sigma^{(2)}_{hh}$ exhibits a
pronounced maximum in the vicinity of the Fermi energy due to the high
probability amplitude for particle-hole excitations near $\varepsilon_F$.
It is in very good agreement with the results obtained
from Brueckner-Hartree-Fock (BHF) calculation with $G$-matrix \cite{ZUO}. 

\begin{figure}[htb]
\includegraphics[width=7cm]{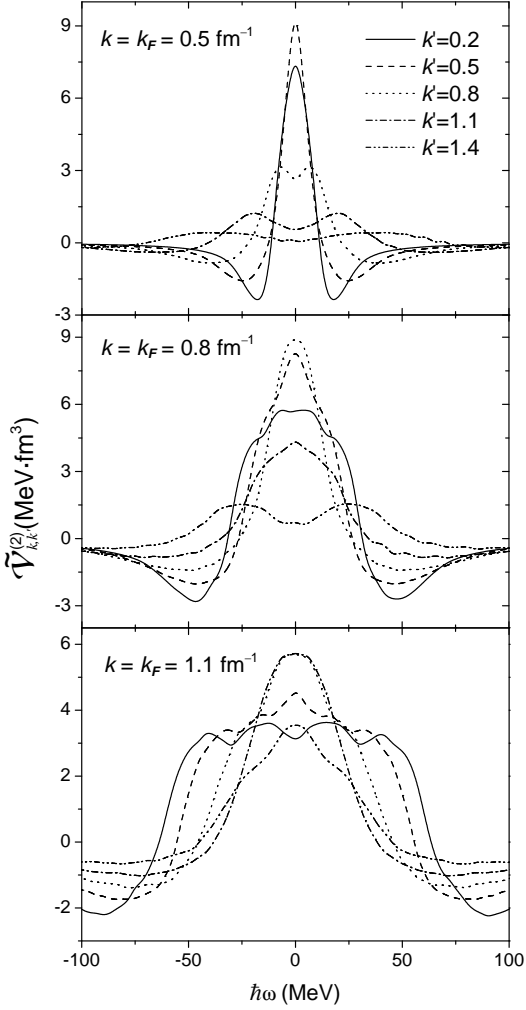}
\caption{\label{fig3} Screening potential vs energy at $k_F$ = 0.5, 0.8, 1.1
         fm$^{-1}$, separately.}
\end{figure} 
 
The second-order potential is given by the one-bubble exchange term (plotted in 
Fig. 1(b), which is the first one of the ring diagram series. Physically it represents
the screening to the pairing due to the medium polarization. 
Again we take the Gogny force for all vertices in Fig. 1(a,b). Our prediction for 
${\cal V}^{(2)}_{kk'}(\omega)$ at three typical densities is reported in Fig. 3.
We plot the symmetric part
\begin{equation}
{\tilde{\cal V}}_{k,k'}^{(2)}(\omega,\omega'_k) \equiv \frac{1}{2(2\pi)^3} [{\cal V}(\omega,\omega'_k)
+{\cal V}(\omega,-\omega'_k)],                          
\end{equation}   
which is the only one relevant for the pairing gap.  
The  strength of $\tilde{\cal V}$ is concentrated around the Fermi energy ($\omega=0$) with a
peak value at $k=k'=k_F$ and a width increasing with the density.
Its $\omega$ dependence is shaped by the polarization part, i.e., 
Lindhard functions \cite{FETT} which at $\omega=0$ 
(static limit) is repulsive at any momentum and density, but it becomes attractive for 
$|\omega|\gg \varepsilon_F$. One therefore expects a reduction of the gap due to screening.


The gap equation, Eq. (1), is to be coupled to the closure equation for the Green's function
\begin{equation}
\rho = \sum\limits_{k} \int \frac{d\omega}{2\pi i} e^{i\omega 0^+}
G_k(\omega)
\end{equation}
fixing the chemical potential in the superfluid phase. 

The anomalous propagator
\begin{equation}
F_{k}(\omega ) = \frac{\Delta(\omega )}{\left[ \hbar \omega
-\varepsilon _{k}(\omega )\right] \left[ \hbar \omega +\varepsilon
_{k}(-\omega )\right] } , 
\end{equation} 
has two poles, symmetric with respect to the imaginary axis of the complex 
$\omega$-plane \cite{BAL}, which are the roots of the equation
\begin{equation}
\pm\omega_k = \varepsilon_k (\pm\omega_k).
\end{equation}
The quasi-particle energy $\varepsilon_k$  is given by
\begin{equation}
\varepsilon _{k}(\omega )= \Sigma^{-}_{\vec k}(\omega )
+\sqrt{\left[
\varepsilon _{k}^{0} + \Sigma^{+}_{\vec k}(\omega )\right]
^{2}+\Delta_{\vec k}^{2}(\omega)}, 
\end{equation}
where
\begin{equation}
 \Sigma^{\pm}_{\vec k}(\omega)=\frac{1}{2}[\Sigma_{\vec k}(\omega )\pm \Sigma_{\vec k}(-\omega )].
\end{equation} 
The two poles are located close to the real axis on opposite sides of the
imaginary axis.
 Leaving aside a general integration of the gap equation,
we adopt the pole approximation relying on replacing the full propagator by its
pole part:
\begin{eqnarray}
F_k(\omega) =\frac{Z_k\Delta_k(\omega)}{\varepsilon_k(\omega)+\varepsilon_k(-\omega)} 
[\frac{1}{\omega-\omega_k+i\eta}-
\frac{1
}{\omega+\omega_k-i\eta}]
\end{eqnarray}
In general the residue $Z_k$ at the poles is defined as
\begin{equation}
Z_{k}=\left( 1-\left. \frac{\partial \varepsilon _{k}
(\omega)}{\partial \omega }\right\vert _{{\omega}
=\omega _{k}}\right) ^{-1},
\end{equation}
calculated  at the Fermi surface. In the calculations we took the limit of $Z_{k}$ for 
$\Delta \rightarrow 0$, which corresponds to the quasi-particle strength. 
The factor $Z$ is keeping the full dynamical dependence of the self-energy
reported in Fig. 2. Afterwards we may subtract the static part $\Sigma^{(2)}(\omega)$  from
the self-energy. In the calculation of the gap the inclusion of the latter
only brings a variation of less than one percent since the gap is not sensitive to
the static self-energy far from $\epsilon_F$.  
Inserting Eq. (8) for the anomalous propagator into the gap equation, after
$\omega$ integration we obtain
\begin{equation}
\Delta _k(\omega)=-\frac{1}{2}\int {k^{\prime 2}dk^{\prime }}{\tilde{\cal V}}_{k,k^{\prime }}
(\omega -\omega _{k^{\prime }})
\frac{Z_{k^{\prime }}\Delta _{k^{\prime}}(\omega_{k^{\prime}})}{\varepsilon_k(\omega_{k'})
+\varepsilon_k(-\omega
_{k'})}.
\label{sec7-4}
\end{equation}

Notice that the reason why only the $\omega$-even part of the interaction contributes to 
the integral can be traced to time reversal invariance of the superfluid ground state for 
which  the anomalous propagator as well as the gap function are  even functions of  $\omega$.

\begin{figure}[htb]
\includegraphics[width=7cm]{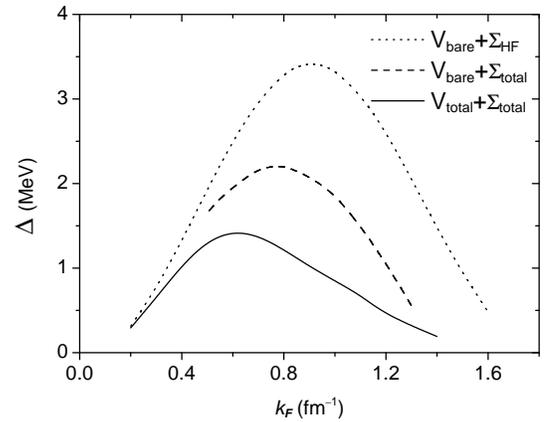}
\caption{\label{fig4} Energy gap in the present approximation. For comparison 
the prediction from the pure BCS model (dotted line) and from BCS plus self-energy
effects (dashed line) are plotted.}
\end{figure} 

The remarkable advantage of this approximation is that the gap depends only 
parametrically on $\omega$ and its energy dependence is only related to
the energy dependence of the interaction.  
The on-shell gap $\Delta_k(\omega_k)$ fulfills the equation
\begin{equation}
\Delta _k(\omega_k)=-\frac{1}{2}\int {k^{\prime 2}dk^{\prime }}
{\tilde{\cal V}}_{k,k^{\prime }}^S(\omega_k -\omega _{k^{\prime }})
\frac{Z_{k^{\prime}}\Delta _{k^{\prime}}(\omega_{k^{\prime}})}{\varepsilon_k(\omega_{k'})
+\varepsilon_k(-\omega
_{k'})}.
\label{sec7-2}
\end{equation}
equivalent to the gap equation in the static limit. 

The approximate version of the gap equation, Eq. (10), has been solved using as
input the self-energy and pairing potential discussed in the previous section.
The focus was in the $^1S_0$ neutron-neutron pairing for neutron   matter, which is 
by far the most important component of pairing as to possible implications in nuclear 
systems. The energy gap at the Fermi surface ($k=k_F$ and $\omega=0$) as a function 
of $k_F$ is reported in Fig. 4. The domain of existence
of the superfluid state is mainly at low densities with a peak value of about 1.4 MeV
at $k_F=0.6$ fm$^{-1}$. In the same figure we also report  the result in the BCS limit 
[for a review see \cite{SCHUL}] (neither self-energy effects nor screening) and the 
result with self-energy effects but without screening \cite{ZUO}. From the 
comparison of the three predictions one sees that the main suppression of $\Delta$
is due to the strong g.s. correlations which lead to a $Z$-factor much less than unit.
However screening of the pairing interaction produces an additional suppression. 
It has to be noticed that the screening potential also shifts the peak value of the gap 
to lower density, where the suppression is less sizeable. This implies that the
medium effects are not expected to reduce dramatically the pairing at the nuclear
surface and the feature of pairing as surface effect is still confirmed by 
our fully consistent calculation of pairing in neutron matter. 
 As final result, obtained solving Eq. (10), we report in Fig. 5 the gap as a 
 function of the energy for $k_F=0.6$ fm$^{-1}$. We address 
 here the relevance of such a feature
for the study of pair correlations in dynamical processes such as the 
expanding and disassembling phases of the fragmentation events in heavy-ion collisions
\cite{BALLOM}. 

\begin{figure}[htb]
\includegraphics[width=7cm]{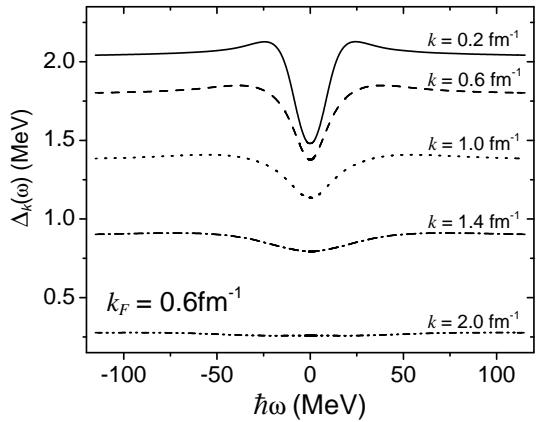}
\caption{\label{fig5} Energy gap as a function of $\omega$ at $k_F=0.6$ fm$^{-1}$. }
\end{figure}

Screening effects on the pairing interaction have also been studied in different
contexts. One of the earliest calculations has been performed in the
framework of the second-order correlated basis perturbation 
theory \cite{CLARK}, where a pairing suppression by a factor 4 is predicted. In that 
work the momentum dependence is mostly neglected
that amounts to overestimate the suppression.
Closer to the present approach is the polarization potential 
\cite{WAMB,CTLI} calculated from the induced interaction theory \cite{BROW}, which gives
substantially the same gap when treated within the Landau parameter approximation \cite{LOMBA}. 
Finally,we should mention a parallel study on finite nuclei \cite{BORT}, where the 
polarization potential is given by the coupling to surface vibrations. While the 
self-energy plays the same role as in neutron matter, the phonon exchange produces 
an enhancement of the pairing 
at variance with the dominant repulsive effect of the spin fluctuations
in neutron matter. 
 
In conclusion, this work constitutes a continuation of a previous one(Ref. \cite{ZUO}) 
where we investigated self-energy effects on the pairing gap in infinite neutron matter.
Here we treated consistently the additional inclusion of vertex corrections. On the same 
footing we considered the lowest order particle hole polarization bubble both in the 
self-energy and in the screening of the pairing force. Instead of the G-matrix we 
used the phenomenological Gogny force at all coupling vertices. We verified that 
this replacement has only very little influence on the numerical results. The 
screened pairing interaction is in principle energy dependent but in the quasi-particle 
approximation used here, this dependence on energy becomes only a parametrical one
which greatly facilitates the numerical task of solving the gap equation. The 
outcome of the inclusion of vertex corrections is that the gap as a function of 
$k_F$ maintains approximately its bell-shaped form, but with respect to the 
self-energy corrections only, a further substantial reduction of the gap is 
induced by screening the bare NN interaction in the gap equation. Therefore 
with respect to the lowest order approach, i.e., without any polarization 
effects, that is with bare interaction and k-mass only, the gap at its maximum 
is now reduced by about 50\%. This strong reduction is a common feature of all 
previous calculations and in this sense our investigation is a confirmation of 
what has been found by other authors earlier, even though the approaches differ
in detail. Nonetheless this strong reduction of the pairing due to the polarization 
remains intriguing. If the same situation should prevail in nuclear matter 
an estimate via the local density approximation \cite{RING} would lead to a 
by far too low value of the gap in finite nuclei. However the influence of 
polarization terms in nuclear matter, due to the different quantum numbers 
involved, may be quite different from the one in neutron matter. It may be an 
extremely interesting problem for studies in the near future to see whether 
the polarization corrections in the different channels, i.e., n-n pairing in 
neutron matter, n-n pairing in nuclear matter and n-p pairing in nuclear 
matter, can, at least qualitatively, explain the 
experimental finding known from finite nuclei.


\end{document}